\documentclass[10pt,letterpaper]{article}
\usepackage[top=0.85in,left=2.75in,footskip=0.75in]{geometry}

\usepackage{bm}
\usepackage{amsmath, amsthm, amssymb}
\usepackage{graphicx}
\usepackage{mathrsfs}
\usepackage{multirow}

\usepackage{fixltx2e}

\usepackage{changepage}

\usepackage[utf8x]{inputenc}

\usepackage{textcomp,marvosym}

\usepackage{cite}

\usepackage{nameref,hyperref}

\usepackage[right]{lineno}

\usepackage{microtype}
\DisableLigatures[f]{encoding = *, family = * }

\usepackage[table]{xcolor}

\usepackage{array}

\newcolumntype{+}{!{\vrule width 2pt}}

\newlength\savedwidth

\raggedright
\usepackage[aboveskip=1pt,labelfont=bf,labelsep=period,justification=raggedright,singlelinecheck=off]{caption}

\makeatletter
\renewcommand{\@biblabel}[1]{\quad#1.}
\makeatother

\date{}

\usepackage{lastpage,fancyhdr,graphicx}
\usepackage{epstopdf}
\fancyheadoffset[L]{2.25in}
\fancyfootoffset[L]{2.25in}
\newcommand{\Part}[3]{ \frac{ \partial^{#3} #1 }{ \partial #2^{#3} } }
\newcommand{\V}[1]{\bm{#1} } 

\newcommand{\mR}{\mathbb{R}}

\newcommand{\lb}{\left(}
\newcommand{\rb}{\right)}
\newcommand{\lbb}{\left\{}
\newcommand{\rbb}{\right\}}

\newcommand{\T}[1]{\tilde{#1}}
\newcommand{\Req}[1]{Eq\ (\ref{eq:#1})}

\newcommand{\NReq}[1]{(\ref{eq:#1})}
\newcommand{\Reqs}[2]{Eqs\ (\ref{eq:#1},\ref{eq:#2})}
\newcommand{\BReqs}[2]{Eqs\ (\ref{eq:#1}) and (\ref{eq:#2})}   

\newcommand{\Rfig}[1]{Fig\ \ref{fig:#1}}

\newcommand{\Lfig}[1]{\label{fig:#1}}
\newcommand{\Leq}[1]{\label{eq:#1}}

\newcommand{\Lsec}[1]{\label{sec:#1}}
\newcommand{\be}{\begin{eqnarray}}
\newcommand{\ee}{\end{eqnarray}}
\newcommand{\ba}{\begin{array}}
\newcommand{\ea}{\end{array}}
\newcommand{\no}{\nonumber}

\newcommand{\subbe}{\begin{subequations}}
\newcommand{\subee}{\end{subequations}}
\newcommand{\Bs}{\backslash}

\newcommand{\argmin}{\mathop{\rm arg~min}\limits}

\newcommand{\Span}[1]{\mathrm{span}\lb #1 \rb }
\newcommand{\lA}{\leftarrow}

\newcommand{\Hessian}{\partial_{\V{x}}^2}

\begin{document} 
\vspace*{0.2in}

\begin{flushleft}
{\Large
\textbf\newline{
Accelerating Cross-validation with Total Variation and Its Application to Super-resolution Imaging} 
}
\newline
\\
Tomoyuki Obuchi\textsuperscript{1*},
Shiro Ikeda\textsuperscript{2},
Kazunori Akiyama\textsuperscript{3,4,5},
Yoshiyuki Kabashima\textsuperscript{1}
\\
\bigskip
\textbf{1} Department of Mathematical and Computing Science/Tokyo Institute of Technology, Yokohama 226-8502, Japan \\
\textbf{2} The Institute of Statistical Mathematics, Tachikawa, Tokyo, 190-8562, Japan \\
\textbf{3} Haystack Observatory/Massachusetts Institute of Technology, Westford, MA 01886, USA \\
\textbf{4} National Astronomy Observatory of Japan, Osawa 2-21-1, Mitaka, Tokyo 181-8588, Japan \\
\textbf{5} Black Hole Initiative, Harvard University, Cambridge, MA 02138, USA
\bigskip

%
%





* obuchi@c.titech.ac.jp
\end{flushleft}


\section*{Abstract}
We develop an approximation formula for the cross-validation error (CVE) of a sparse linear regression penalized by $\ell_1$-norm and total variation terms, which is based on a perturbative expansion utilizing the largeness of both the data dimensionality and the model. The developed formula allows us to reduce the necessary computational cost of the CVE evaluation significantly. The practicality of the formula is tested through application to simulated black-hole image reconstruction on the event-horizon scale with super resolution. The results demonstrate that our approximation reproduces the CVE values obtained via literally conducted cross-validation with reasonably good precision.



\section{Introduction}
At present, in many practical situations of science and technology, large high-dimensional observational datasets are created and accumulated on a continuous basis. An essential difficulty concerning the treatment of such high-dimensional data is the extraction of meaningful information. Sparse modeling~\cite{Rish:14,Hastie:15} is a promising framework for overcoming this difficulty, which has recently been utilized in many disciplines~\cite{Okada:13,Mairal:14}. In this framework, a statistical or machine-learning model with a large number of parameters (explanatory variables) is fitted to the data, in conjunction with a certain sparsity-inducing penalty. This penalty should be appropriately chosen with consideration of the processed data. One representative penalty is the $\ell_1$ regularization, which retains certain preferred properties, such as the statistical model convexity~\cite{Tibshirani:96,Efron:04}. A similar penalty that has received more recent focus is the so-called ``total variation (TV)''~\cite{Rudin:92,Chambolle:04,Beck:09-1}, which can be regarded as the $\ell_1$ regularization imposed on the difference between neighboring explanatory variables. The TV yields ``continuity'' of the neighboring variables, which is suitable for the processing of certain datasets expected to have such continuity, such as natural images~\cite{Mairal:14,Rudin:92,Chambolle:04,Beck:09-1}.

Another common difficulty associated with the use of statistical models is model selection. In the context of image processing using the $\ell_1$ and TV regularizations, this difficulty appears during the selection of appropriate regularization weights. A practical framework to select these weights, which is applicable to general situations, is cross-validation (CV). CV provides an estimator of the statistical-model generalization error, {\it i.e.}, the CV error (CVE), using the data under control, and the minimum CVE obtained when sweeping the weights yields the optimal weight values. This versatile framework is, however, computationally demanding for large datasets/models, and this problem frequently becomes a bottleneck affecting model selection. Thus, reducing the CVE computational cost could have a significant impact on a broad range of sparse modeling applications in various disciplines. 

Considering these circumstances, in this paper, we provide a CVE approximation formula for a statistical model of linear regression penalized by the $\ell_1$ and TV terms, to efficiently reduce the computational cost. Note that the formula for the case penalized by the $\ell_1$ term alone has already been proposed in~\cite{Obuchi:16}, and the formula presented herein is a generalization of it. Below, we show the formula derivation and perform a demonstration in the context of super-resolution imaging. The processed images employed in this study are reconstructed from simulated observations of black holes on the event-horizon scale for the Event Horizon Telescope (EHT, see \cite{EHT,Asada:17,Akiyama:15}) full array. Note that our formula will be applied to actual EHT observations to be conducted after April 2017. 

\section{Problem setting}
Let us suppose that our measurement is a linear process, and denote the measurement result as $\V{y}\in \mR^{M}$ and the measurement matrix  as $A=\lbb A_{\mu i} \rbb_{\mu=1,\cdots,M;\, i=1,\cdots N}  \in \mR^{M\times N}$. The explanatory variables, corresponding to the images that will be examined in the later demonstration, are denoted by $\V{x} \in \mR^{N}$. The quality of the fit to the data is described by the residual sum of squares (RSS), {\it i.e.}, $\mathcal{E}(\V{x}|\V{y},A)=\frac{1}{2}|| \V{y}-A\V{x} ||_{2}^2$. In addition, we consider the following penalty consisting of $\ell_1$ and TV terms:
\be
R(\V{x};\lambda_{\ell_1},\lambda_{T})=
\lambda_{\ell_1}||\V{x}||_{1}+\lambda_{T}T(\V{x}),
\Leq{R}
\ee
where the $T(\V{x})$ term corresponds to the TV and is expressed as 
\be
T(\V{x})=\sum_{i}\sqrt{\sum_{j \in \partial i}(x_j-x_i)^2 }
\equiv \sum_{i}t_{i}
,
\Leq{TV}
\ee
and $\partial i$ denotes the neighboring variables of the $i$th variable. There is some variation in the definition of ``neighbors''; here, we follow the standard approach~\cite{Rudin:92,Chambolle:04,Beck:09-1}. That is, $\V{x}$ is assumed to be a two-dimensional image and the neighbors of the $i$th pixel correspond to the right and down pixels. However, the bottom row (the rightmost column) of the image is exceptional, as the neighbor of each pixel in that row (column) corresponds to the right (down) pixel only. Note that the developed approximation formula presented below is independent of this specific choice of neighbors and can be applied to general cases.

For this setup, we consider the following linear regression problem with the penalty given in \Req{R}
\be
\hat{\V{x}}(\lambda_{\ell_1},\lambda_{T})=
\argmin_{\V{x}}
\lbb 
\mathcal{E}(\V{x}|\V{y},A) +R(\V{x};\lambda_{\ell_1},\lambda_{T})
\rbb, 
\Leq{Problem}
\ee
where $\argmin_{u}\{f(u)\}$ generally represents the argument that minimizes 
an arbitrary function $f(u)$. 
Further, we consider the leave-one-out (LOO) CV of \Req{Problem} in the form
\be
&&
\hat{\V{x}}^{\Bs \mu}(\lambda_{\ell_1},\lambda_{T})
=
\argmin_{\V{x}}
\lbb 
\frac{1}{2} \sum_{\nu (\neq \mu)}\lb y_{\nu}-A_{\nu i}x_i \rb^2 +R
\rbb
\no \\ &&
\equiv
\argmin_{\V{x}}
\lbb 
\mathcal{E}(\V{x}|\V{y}^{\Bs \mu},A^{\Bs \mu})  +R(\V{x};\lambda_{\ell_1},\lambda_{T})
\Leq{LOOProblem}
\rbb.
\ee
Note that the system without the $\mu$th row of $\V{y}$ and $A$ is referred to as the ``$\mu$th LOO system'' hereafter. 
In this procedure, the CVE, {\it i.e.}, the generalization error estimator, is 
\be
\mathcal{E}_{\rm LOO}(\lambda_{\ell_1},\lambda_{T})
=
\frac{1}{2}\sum_{\mu=1}^{M}( y_{\mu} - \V{a}_{\mu}^{ \mathrm{T} } \hat{\V{x}}^{\Bs \mu}(\lambda_{\ell_1},\lambda_{T})
)^2,
\Leq{LOOE}
\ee
where $\V{a}_{\mu}^\top =(A_{\mu 1},\cdots,A_{\mu N})$ is the $\mu$th row vector of $A$. We term this simply the ``LOO error (LOOE).''

Computing the LOOE requires solution of \Req{LOOProblem} $M$ times, by definition, which is computationally expensive. Therefore, the purpose of this paper is to avoid this computational expense by deriving an approximation formula of \Req{LOOE}. 

\section{Approximation formula for softened system}
When $M$ is sufficiently large, {\it i.e.}, the number of observations is large enourgh, the difference between the LOO solution $\hat{\V{x}}^{\Bs \mu}$ and the full solution $\hat{\V{x}}$ is expected to be small. This intuition naturally motivates us to conduct a perturbation connecting these two solutions. To conduct this perturbation, we ``soften'' the penalty by introducing a small cutoff $\delta(>0)$ in the TV, having the form 
\be
R \to R^{\delta}(\V{x};\lambda_{\ell_1},\lambda_{T})
=\lambda_{\ell_1}\sum_{i}||\V{x}||+\lambda_{T}T^{\delta}(\V{x}),
\ee
where
\be
T^{\delta}
=\sum_{i}\sqrt{\sum_{j \in \partial i}(x_j-x_i)^2 +\delta^2}
\equiv \sum_{i}t^{\delta}_{i}
\Leq{cutoffTV}
.
\ee
An approximation formula in the presence of $\ell_1$ regularization with smooth cost functions has already been proposed in~\cite{Obuchi:16}. We employ that formula here and take the limit $\delta \to 0$. 

To state the approximation formula, we begin by defining ``active''  and ``killed'' variables. Owing to the $\ell_1$ term, some variables are set to zero in $\hat{\V{x}}$; we refer to these variables as ``killed variables.'' The remaining finite variables are termed ``active variables.'' We denote the index sets of the active and killed variables by $S_A$ and $S_K$, respectively. The active (killed) components of a vector $\V{x}$ are formally expressed as $\V{x}_{S_A}(\V{x}_{S_K})$. For any matrix $X$, we use double subscripts in the same manner. For example, for an $N\times N$ matrix, a submatrix having row and column components of $S_A$ and $S_K$, respectively, is denoted by $X_{S_A S_K}$. 

The approximation formula can be derived through the following two steps. Note that, in this derivation, a crucial assumption is that the sets of active and killed variables are common among the full and LOO systems. This assumption may not 
hold exactly in practice, but the resultant formula is asymptotically exact in the large-$N$ limit~\cite{Obuchi:16}.

The first step is to compute the values of the active variables and their response to small perturbation. The active variables are determined by the extremization condition of the softened cost function with respect to the active variables, such that
\be
\Part{\lb \mathcal{E}(\V{x}|\V{y}^{\Bs \mu},A^{\Bs \mu} )+R^{\delta}(\V{x};\lambda_{\ell_1},\lambda_{T})\rb}{\V{x}_{S_A}}{} 
=\V{0}
\Rightarrow (\hat{\V{x}}^{\delta \Bs \mu})_{S_A}.
\ee
The focus here is the response of this solution when a small perturbation $-\V{h}\cdot \V{x}$ is incorporated into the cost function. A simple computation demonstrates that the active--active components of the response function, $\lb \chi^{\delta \Bs \mu} \rb_{S_A S_A} =\left. \Part{}{\V{h}_{S_A} }{}(\hat{\V{x}}^{\delta \Bs \mu})_{S_A}\right|_{\V{h}=0}$, are equivalent to the inverse of the cost-function Hessian
\be
&&
\lb \chi^{\delta \Bs \mu} \rb_{S_A S_A}=\lb H^{\delta \Bs \mu}_{S_A S_A} \rb^{-1},
\\ &&
H^{\delta \Bs \mu}=\Hessian \lb \mathcal{E}(\V{x}|\V{y}^{\Bs \mu},A^{\Bs \mu} )+R^{\delta}(\V{x};\lambda_{\ell_1},\lambda_{T})\rb
=G^{\Bs \mu}+\Hessian R^{\delta}(\V{x};\lambda_{\ell_1},\lambda_{T}) ,
\ee
where $\Hessian$ denotes the Hessian operator $\Hessian\equiv (\frac{\partial^2}{\partial x_i \partial x_j})$ and $G^{\Bs \mu}$ is the Gram matrix of $A^{\Bs \mu}$, {\it i.e.}, $G^{\Bs \mu} \equiv \lb A^{\Bs \mu} \rb^\top A^{\Bs \mu}$. The other components of the response function are identically zero, from the stability assumption of $S_K$ and because the killed variables are zero, with $\hat{\V{x}}_{S_K}=\hat{\V{x}}_{S_K}^{\Bs \mu}=\V{0}$.

In the second step, we connect the full solution to the LOO solution, through the above perturbation with an appropriate $\V{h}$. To specify the perturbation, we assume that the difference $\V{d}^{\delta}=\hat{\V{x}}^{\delta} -\hat{\V{x}}^{\delta \Bs \mu}$ is small and expand the RSS of the full system with respect to $\V{d}^{\delta}$ as follows:
\be
&&
\hspace{-1cm}
\mathcal{E}(\hat{\V{x}}^{\delta \Bs \mu}|\V{y},A) 
\!\approx \!\mathcal{E}(\hat{\V{x}}^{\delta}|\V{y},A)
\!- \!\sum_{\mu=1}^{M}(y_{\mu}\! - \!\V{a}_{\mu}^\top\hat{\V{x}}^{\delta} )\V{a}_{\mu}^\top \V{d}^{\delta}.
\ee
This equation implies that the perturbation between the full and LOO systems can be expressed as $\V{h}_{\mu}=(y_{\mu}-\V{a}_{\mu}^\top \hat{\V{x}}^{\delta})\V{a}_{\mu}$. Hence, we obtain
\be
&&
\hspace{-1cm}
\hat{\V{x}}^{\delta}
\!\approx \! \hat{ \V{x} }^{\delta \Bs \mu} \!+ \!\chi^{\delta \Bs \mu} \V{h}_{\mu}
\!=\!
\hat{\V{x}}^{\delta \Bs \mu}
\!+\!
(y_{\mu}\!-\!\V{a}_{\mu}^\top\hat{\V{x}}^{\delta})
\chi^{\delta \Bs \mu} \V{a}_{\mu }.
\Leq{x-perturbed}
\ee
The Hessian of the full system has a simple relationship with the LOO Hessian, such that
\be
&&
H^{\delta}
\equiv 
G^{\Bs \mu}+(\V{a}_{\mu} \V{a}^\top _{\mu})+\Hessian R^{\delta}(\hat{\V{x}}^{\delta} )
\approx
H^{\delta \Bs \mu}+\lb \V{a}_{\mu}\V{a}^\top _{\mu} \rb,
\Leq{responses}
\ee
where the approximation at the righthand side comes from replacing $\hat{\V{x}}^{\delta}$ with $\hat{\V{x}}^{\delta \Bs \mu}$ in the argument of $R^{\delta}(\V{x})$. Inserting \Reqs{x-perturbed}{responses} in conjunction with $\chi^{\delta}_{S_A S_A}=(H^{\delta}_{S_A S_A})^{-1}$ into \Req{LOOE} and using the Sherman-Morrison formula for matrix inversion, we find
\be
\mathcal{E}_{\rm LOO}(\lambda_{\ell_1},\lambda_{T})
\!\approx \!
\frac{1}{2} \!
\sum_{\mu=1}^{M}
\frac{
(y_{\mu}-\V{a}_{\mu}^\top\hat{\V{x}}^{\delta})^2
}
{
\lb \! 1\! - \!\V{a}_{\mu S_A}^\top \!\lb \chi^{\delta} \rb_{S_A S_A} \! \V{a}_{\mu S_A}  \! \rb^{2}
}
.
\Leq{LOOE-approx}
\ee
According to \Req{LOOE-approx}, we can compute the LOOE only from the full solution $\hat{\V{x}}^{\delta}$, without
actually performing CV, which facilitates considerable reduction of the computational cost. 

\section{Handling a singularity}
Let us generalize \Req{LOOE-approx} to the limit $\delta \to 0$, where the penalty contains another singular term in addition to the $\ell_1$ term. This TV singularity tends to ``lock'' some of the neighboring variables, {\it i.e.}, $x_{j}=x_{i}~(\forall j \in \partial i)$, which corresponds to $t_i=0$ in \Req{TV}. If two different vanishing TV terms, $t_i$ and $t_j$, share a common variable $x_r$, all the variables in those TV terms take the same value $x_{k}=x_{r}~(\forall{k} \in (\{i\} \cup \partial i \cup \{j\} \cup \partial j) )$. In this manner, the active variables are separated into several ``locked'' clusters, with all the variables inside a cluster having an identical value. This implies that the variable response to a perturbation, $\chi=\lim_{\delta \to 0}\chi^{\delta}$, should have the same value for all variables in a cluster and may, therefore, be merged. Below, we demonstrate this behavior for the $\delta \to 0$ limit. For the derivation, we assume that the clusters are common to both  the full and LOO systems, similar to the assumption for $S_A$ and $S_K$.  For convenience, we index the clusters by $\alpha,\beta \in C$ and denote the number of clusters by $|C|$; the index set of variables in a cluster $\alpha$ is represented by $S_{\alpha}$ and the total set of indices in all  clusters is denoted by $S_{C}\equiv \cup_{\alpha}S_{\alpha}$. Hereafter, we concentrate on the active variable space only and omit the killed variable space. The complement set of $S_{C}$, {\it i.e.}, the set of isolated variables that do not belong to any cluster, is denoted by $S_{I}$ and, thus, $S_{A}=S_{I}\cup S_{C}$. 

Two crucial observations for the derivation are the ``scale separation'' and the presence of the ``zero mode.'' For vanishing TV terms, a natural scaling to satisfy $\lim_{\delta \to 0} t_{i}^{\delta}=t_i=0$ is $|\hat{x}^{\delta}_j-\hat{x}^{\delta}_i|\propto \delta ~(\forall{j}\in \partial i)$. Once this scaling is assumed, we realize that the components of the Hessian that are directly related to the clusters diverge. 
Let us define by $\hat{S}_{\alpha}$ the set of TV terms corresponding to cluster $\alpha$, {\it i.e.}, $\hat{S}_{\alpha}=\{i| \lb \{i\} \cup \partial i \rb \subset S_{\alpha} \} $. Hence, by construction and for all $\alpha \in C$, all components of $D_{\alpha}^{\delta}\equiv \lambda_T \lb \Hessian  \sum_{i\in \hat{S}_{\alpha} } t_i^{\delta}  \rb_{S_{\alpha} S_{\alpha}}$ are scaled as $1/\delta$ and, thus, diverge as $\delta \to 0$. The remaining terms are retained as $O(1)$. According to this ``scale separation,'' we decompose the Hessian as $H^{\delta}=D^{\delta}+F^{\delta}$, where $D^{\delta}$ is the direct sum of the diverging components in the naively extended space; $D^{\delta}=\bigoplus_{\alpha} D^{\delta}_{\alpha}$; and $F^{\delta}$ consists of the remaining $O(1)$ terms. This decomposition can be schematically expressed as 
\be
&&
H^{\delta}=D^{\delta}+F^{\delta}
= \!\!
 \left( \!\!
 \begin{array}{c|c}
 \begin{array}{c|c|c}
D^{\delta}_{1} & 0 & 0 \\
\hline
  0 & \ddots & 0 \\
\hline
0 & 0 & D^{\delta}_{|C|} \\
 \end{array}
 & 0 \\
\hline
0 & 0
\end{array}
\!\! \right)
 \!\!+\!\!
  \left( \!\!\!
 \begin{array}{c|c}
 \begin{array}{ccc}
 &  &  \\
  & F^{\delta}_{S_C S_C} &  \\
 &  & \\
 \end{array}
 & F^{\delta}_{S_C S_I} \\
\hline
F^{\delta}_{S_I S_C} & F^{\delta}_{S_I S_I}
\end{array}
 \!\!\! \right). 
 \Leq{separation}
\ee

We denote the basis of the current expression by $\{ \V{e}_{i} \}_{i\in S_{A}}$, with $(\V{e}_{i})_j=\delta_{ij}$, and move to another basis that diagonalizes $D^{\delta}_{S_C S_C}$. Each $D_{\alpha}^{\delta}$ has a ``zero mode,'' and its normalized eigenvector is given by $\V{z}_{\alpha}=(z_{i \alpha})$, where $z_{i \alpha} =1/\sqrt{|S_{\alpha}|}$ for $i\in S_{\alpha}$ and $0$ otherwise, in the full space. This behavior originates from the symmetry, such that the $\{t_{i}^{\delta}\}_{i \in \hat{S}_{\alpha}} $ are invariant under a uniform shift in the $S_{\alpha}$ sub-space, {\it i.e.}, $x_j \to x_j + \Delta$ $(\forall j \in S_{\alpha})$ for $\forall{\Delta} \in \mathbb{R}$. This invariance can also be directly seen from a property of the Hessian, {\it i.e.}, $\frac{\partial^2}{\partial x_i^2}t_i^{\delta}+\sum_{j \in \partial i} \frac{\partial}{\partial x_i \partial x_j}t_i^{\delta}=0$.

In addition, we represent the set of normalized eigenvectors of all the other modes of $D_{\alpha}^{\delta}$, 
which have eigenvalues $\lambda_{\alpha a}$ that are proportional to $1/\delta$ and positively divergent, 
as $\{ \V{u}_{\alpha a} \}_{a=1}^{|S_{\alpha}|-1}$. 
Then, $\{\{ \{ \V{u}_{\alpha a} \}_{a},\V{z}_{\alpha} \}_{\alpha}\}$
diagonalizes $D^{\delta}_{S_C S_C}$ and 
$\{ \{ \{ \V{u}_{\alpha a} \}_{a},\V{z}_{\alpha} \}_{\alpha}, \{ \V{e}_{i} \}_{i\in S_I} \}$ constitutes 
an orthonormal basis of the full space. 
Corresponding to this variable change, we denote $\T{S}_{Z}$, $\T{S}_{I+Z}$, and $\T{S}_{C-Z}$
as the index set of variables in the space spanned by $\{  \V{z}_{\alpha} \}_{\alpha} $, 
$\{ \{ \V{z}_{\alpha} \}_{\alpha}, \{ \V{e}_{i} \}_{i\in S_I} \}$, and
$\{ \V{u}_{\alpha a} \}_{\alpha,a}$, respectively.  
In the new expression, we can rewrite $H^{\delta}=D^{\delta}+F^{\delta}$ as
\be
&&
H^{\delta}
=
 \left(
 \begin{array}{c|c}
D^{\delta}_{\T{S}_{C-Z}\T{S}_{C-Z}} & 0  \\
\hline
 0 & 0 \\
 \end{array}
 \right) 
+ \left(
 \begin{array}{c|c}
F^{\delta}_{\T{S}_{C-Z}\T{S}_{C-Z}} & F^{\delta}_{\T{S}_{C-Z}\T{S}_{I+Z}}  \\
\hline
F^{\delta}_{\T{S}_{I+Z}\T{S}_{C-Z}} & F^{\delta}_{ \T{S}_{I+Z}\T{S}_{I+Z} }  \\
 \end{array}
 \right),
\ee
where $D^{\delta}_{\T{S}_{C-Z}\T{S}_{C-Z}}=\mathrm{diag} \lb \{ \lambda_{\alpha a}\}_{\alpha,a} \rb$. Because of the divergence of $D^{\delta}_{\T{S}_{C-Z}\T{S}_{C-Z}}$, only $F^{\delta}_{ \T{S}_{I+Z}\T{S}_{I+Z} }$ is relevant for the evaluation of $(H^{\delta})^{-1}$. These considerations yield the explicit formula of $\chi$ as
\be
&&
\hspace{-1cm}
(H^{\delta})^{-1}
 \!
 =
 \!
  \left( \!\!
 \begin{array}{c|c}
(D^{\delta}_{\T{S}_{C-Z}\T{S}_{C-Z}})^{-1} & 0\\
\hline
0 & \lb F^{\delta}_{ \T{S}_{I+Z}\T{S}_{I+Z} } \rb^{-1} \\
 \end{array}
 \!\! \right)
 \! +  \! O(\delta)
\no \\  &&
\hspace{-1cm}
\to 
 \left(
 \begin{array}{c|c}
0 & 0  \\
\hline
0 & \lb F_{ \T{S}_{I+Z}\T{S}_{I+Z} } \rb^{-1} \\
 \end{array}
 \right)=\chi,
\Leq{chi1}
\ee
where $F=\lim_{\delta \to 0}F^{\delta}$. 

By construction, in the reduced space to $\Span{ \{ \V{z}_{\alpha} \}_{\alpha} , \{ \V{e}_{i} \}_{i\in S_{I}} }$, $F_{ \T{S}_{I+Z}\T{S}_{I+Z} }$ can be expressed as  
\be
&&
F_{ \T{S}_{I+Z}\T{S}_{I+Z} }=
\sum_{\alpha,\beta}
\lb 
F_{\alpha \beta}\V{z}_{\alpha}\V{z}_{\beta}^\top 
+
F_{\beta \alpha }\V{z}_{\beta}\V{z}_{\alpha}^\top 
\rb
\no \\ &&
+
\sum_{\alpha}\sum_{i \in S_{I}}\lb F_{\alpha i} \V{z}_{\alpha}\V{e}_{i}^\top+F_{i\alpha }\V{e}_{i}\V{z}_{\alpha}^\top \rb
+\sum_{i,j \in S_{I}}F_{ij}\V{e}_{i}\V{e}_{j}^\top .
\Leq{F_{I+Z}}
\ee
As the non-zero components of the zero mode $\V{z}_{\alpha}$ are
identically given as $1/\sqrt{|S_{\alpha}|}$, all these coefficients can be easily expressed by the original coefficients $F_{ij}$, as
\subbe
\Leq{F-components}
\be
&&
F_{\alpha \beta}=\V{z}_{\alpha}^\top F\V{z}_{\beta}
=\frac{1}{\sqrt{|S_{\alpha}||S_{\beta}|}} \sum_{i\in S_{\alpha},j\in S_{\beta}}F_{ij},
\\ &&
F_{\alpha i}=\V{z}_{\alpha}^\top F\V{e}_{i}=
\frac{1}{\sqrt{|S_{\alpha}|}}\sum_{j\in S_{\alpha}}F_{ji},
\ee
\subee
and $F_{i\alpha}=F_{\alpha i}$ by the symmetry. Now, all the components are explicitly specified. The form of $\chi$ in the original basis $\{ \V{e}_i \}_{i\in S_{A}}$ can be accordingly assessed by moving back from the basis $\{ \{ \{ \V{u}_{\alpha a} \}_{a},\V{z}_{\alpha} \}_{\alpha}, \{ \V{e}_{i} \}_{i\in S_I} \}$, which completes the computation.

Some additional consideration of the above computation demonstrates that we can shorten some steps and obtain a more interpretable result. We introduce a $|\T{S}_{I+Z}|\times |\T{S}_{I+Z}|$ matrix $\bar{F}$ as
\be
\bar{F}_{\alpha \beta}=\sqrt{|S_{\alpha}||S_{\beta}|}F_{\alpha \beta},~
\bar{F}_{\alpha i}=\sqrt{|S_{\alpha}|}F_{\alpha i},
\Leq{Fbar}
\ee
with the remaining components being identical to those of $F_{\T{S}_{I+Z}\T{S}_{I+Z}}$, {\it i.e.}, $\bar{F}_{S_{I}S_{I}}=F_{S_{I}S_{I}}$. \BReqs{F-components}{Fbar} indicate that $\bar{F}$ is simply the matrix summing the rows and columns in each cluster to a row and a column. It is natural that $\bar{F}$ has a direct connection to $\chi$, because the locked variables in a cluster should exhibit the same response against perturbation. In fact, the response function $\chi$ in the original basis is expressed using $\bar{F}$ as
\be
&&
\hspace{-1cm}
\chi
=\sum_{i,j\in S_{I}} \bar{F}^{-1}_{ij} \lb \V{e}_{i}\V{e}_{j}^\top +\V{e}_{j}\V{e}_{i}^\top\rb
+\sum_{\alpha,\beta} \bar{F}_{\alpha \beta}^{-1}\sum_{i\in S_{\alpha}}\sum_{j \in S_{\beta}} \V{e}_{i}\V{e}_{j}^\top 
\no \\ &&
\hspace{-1cm}
+\sum_{\alpha} 
\lb
\sum_{i\in S_{\alpha}}\sum_{j \in S_{I}} \bar{F}^{-1}_{\alpha j} \V{e}_{i}\V{e}_{j}^\top 
+
\sum_{i\in S_{I}}\sum_{j \in S_{\alpha}} \bar{F}^{-1}_{i \alpha} \V{e}_{i}\V{e}_{j}^\top 
\rb.
\Leq{chi2}
\ee
This can be directly shown from \Reqs{chi1}{F-components}, using the relation $F_{\T{S}_{I+Z} \T{S}_{I+Z}}=P\bar{F}_{\T{S}_{I+Z} \T{S}_{I+Z}}P$ with $P=\mathrm{diag}\lb \{ \{1 \}_{i\in S_{I}},\{ \sqrt{|S_\alpha|}^{-1}\}_{\alpha} \} \rb$, and the blockwise matrix inversion formula. \BReqs{LOOE-approx}{chi2} constitute the main result of this paper.  

\section{Algorithmic implementation} 
\subsection{Numerical stability and the softening constant $\delta$}\Lsec{Numerical stability}
For handling the singularity of the cost-function Hessian, we have introduced the softening constant $\delta$ in the TV and finally taken the $\delta \to 0$ limit. In practical implementations, however, we should keep $\delta$ small but finite. To see the reason, it is sufficient to see a simple example with just three variables $\{x_i\}_{i=1}^3$. The softened TV is defined as 
\be
T^{\delta}(\V{x})=\sqrt{(x_2-x_1)^2+(x_3-x_1)^2+\delta^2}=\sqrt{p^2+q^2+\delta^2},
\ee
where $p=x_2-x_1,~q=x_3-x_1$ are introduced. The corresponding gradient and Hessian are 
\be
&&
\Part{ T^{\delta}}{\V{x}}{}=\frac{1}{\lb p^2+q^2+\delta^2\rb^{1/2}} 
\left(
\begin{array}{c}
-p-q    \\
  p \\
  q
\end{array}
\right)
, \\ &&
\Hessian T^{\delta}
=\frac{1}{\lb p^2+q^2+\delta^2\rb^{3/2}} 
\left(
\begin{array}{ccc}
(p-q)^2+2\delta^2  & pq-q^2-\delta^2  & pq-p^2-\delta^2  \\
 pq-q^2-\delta^2    & q^2+\delta^2       & -pq  \\
pq-p^2 -\delta^2    & -pq                      &   p^2+\delta^2
\end{array}
\right).
\ee
The zero point of the gradient is given by $p=q=0$ irrespectively of the $\delta $ value. Inserting this into the Hessian, we get one zero mode proportional to $(1,1,1)^{\top}$ and two finite modes whose eigenvalues are $(3/\delta,~1/\delta)$ being divergent in the $\delta\to 0$ limit. This exactly matches with the assumptions of the approximation formula.

On the other hand, if we first take the limit $\delta\to 0$ before taking the zero gradient limit $p,q\to 0$, we see that two zero modes appear: One is proportional to $(1,1,1)^{\top}$ and the other is to $(p+q,q-2p,p-2q)^{\top}$. This is a bad news because the second zero mode, which remains even in the limit $p,q \to 0$, is never taken into account when deriving the approximation formula: The derivation essentially depends on how the zero mode behaves and our formula loses its justification if such unexpected zero modes exist. 

These considerations manifest that the two limits, $\lim_{\delta \to 0}$ and $\lim_{p,q\to 0}$, are not exchangeable in the TV Hessian. The derivation of our approximation formula assumes $\lim_{\delta \to 0}\lim_{p,q\to 0}$ and thus the algorithmic implementation should reflect this limit in a certain way. A simple way is to keep $\delta$ small but finite, which is actually a common technique to enhance the numerical stability when using the TV~\cite{Chan:01}. The choice of the amplitude of $\delta$ is related to the numerical precision when solving the optimization problem \NReq{Problem}. A practical choice is stated in the next subsection.

\subsection{Procedures}
Here, we state the procedures for implementation of \Reqs{LOOE-approx}{chi2} in a numerical computation. Suppose that we have an algorithm to solve \Req{Problem} and to provide the solution $\hat{\V{x}}$ given $\V{y},A,\lambda_{\ell_1}$, and $\lambda_{T}$. Using this solution and introducing a finite $\delta$ in the Hessian by the reason discussed above, we can assess the LOOE through the following steps:
\begin{enumerate}
 \item{The sets of active and killed variables, $S_A$ and $S_K$, are specified from $\hat{\V{x}}$.}
 \item{The values of all TV terms$\{ t^{\delta}_{i}(\hat{\V{x}}) \}_{i=1}^{N}$ are computed.}
 \item{All clusters $C$ and the index sets belonging to the clusters $\{S_{\alpha} \}_{\alpha \in C}$ are enumerated from $\{ t^{\delta}_{i}(\hat{\V{x}}) \}_{i=1}^{N}$, as well as the one of isolated variables, $S_{I}$.}
 \item{The total variation from which the vanishing TV terms are removed is denoted by $\T{T}^{\delta}(\hat{\V{x}})$, and the regular part of the Hessian is computed as $F=G+\lambda_{T}\Hessian \T{T}^{\delta}(\hat{\V{x}})$.}
 \item{A new index set $S_{R}=\{ \{ \alpha \}_{\alpha \in C}, S_{I}\}$ is defined. }
 \item{On $S_R$, the merged Hessian $\bar{F}$ is constructed from $F$, as $\bar{F}_{S_I S_I}=F_{S_I S_I}$, $\bar{F}_{\alpha \beta}=\sum_{i\in S_{\alpha},i\in S_{\beta}}F_{ij}$,  $\bar{F}_{\alpha S_I}=\sum_{i\in S_{\alpha}}F_{i S_I}$, and   $\bar{F}_{S_I \alpha}=\sum_{i\in S_{\alpha}}F_{S_I i}$. Similarly, the merged measurement matrix $\bar{A}$ is defined as  $\bar{A}_{\mu S_I}=A_{\mu S_I}$, $\bar{A}_{\mu \alpha}=\sum_{i\in S_{\alpha}}A_{\mu i}$. }
 \item{Using $\bar{F}$ and $\bar{A}$, the LOOE factor in \Req{LOOE-approx} is computed as $1-\V{a}^{T}_{\mu S_A}\chi_{S_A S_A}\V{a}_{\mu S_A}=1-\bar{\V{a}}^{T}_{\mu S_R}\lb \bar{F}_{S_R S_R}\Bs \bar{\V{a}}_{\mu S_R}\rb $, where $\bar{\V{a}}^{T}_{\mu}$ is the $\mu$th row vector of $\bar{A}$ and $\V{x}=A \Bs \V{b}$ is the solution of the linear equation $ A\V{x}=\V{b}$.}
 \item{Using the LOOE factor and $\hat{\V{x}}$, the LOOE is evaluated from \Req{LOOE-approx}.}
\end{enumerate}
At step $7$, we take the left division $\bar{F}_{S_R S_R}\Bs \bar{\V{a}}_{\mu S_{R}}$ instead of the inverse $\chi=\bar{F}^{-1}$ for numerical stability. The cluster enumeration at step $3$ involves a delicate point in the definition of $C$ and $\{S_{\alpha} \}_{\alpha \in C}$. Because of the limited precision in the numerics, the TV term $|\hat{x}_j-\hat{x}_i|~(j\in \partial i)$ never exactly vanishes; therefore, we need a certain threshold to extract the cluster structure from the TV terms. Here, we introduce the threshold $\theta$ and enumerate the clusters as follows: 
\begin{enumerate}
\item[3-1.]{If $t^{\delta}_{i}(\hat{\V{x}})\leq \delta+\theta$, the variables in $\{i \} \cup \partial i$ are regarded as ``linked.'' All the links are enumerated by testing $t^{\delta}_{i}(\hat{\V{x}})\leq \delta+\theta$ for all $i=1,\cdots,N$. The set of links is denoted by $L$, and the index set of all variables in $L$ is denoted by $S_{L}$. }
\item[3-2.]{An empty set $C=\phi$ is prepared and the cluster index $\alpha=1$ is defined.}
\item[3-3.]{The following steps are repeatedly implemented while $L$ is non-empty:}
\begin{enumerate}
\item[(i).]{Two empty sets, $S_{\rm tmp}=\phi$ and $S_{\rm cluster}=\phi$, are prepared;}
\item[(ii).]{One link is selected and removed from $L$. The variable indices in the link are entered into $S_{\rm tmp}$;}
\item[(iii).]{The following steps are repeatedly implemented while $S_{\rm tmp}$ is non-empty:}
\begin{enumerate}
\item[a.]{One index $i$ in $S_{\rm tmp}$ is selected and moved from $S_{\rm tmp}$ to $S_{\rm cluster}$;}
\item[b.]{If the above chosen index $i$ exists in $S_{L}$, all the links to $i$ are removed from $L$, and $S_{L}$ is updated accordingly. The variables linked to $i$ are entered into $S_{\rm tmp}$;}
\item[c.]{$S_{\rm tmp} \lA S_{\rm tmp}-S_{\rm cluster}$.}
\end{enumerate}
\item[(iv).]{The variables in $S_{\rm cluster}$ constitute a cluster. $S_{\alpha}=S_{\rm cluster}$ is defined and $\alpha$ is entered into $C$;  } 
\item[(v).]{$\alpha \lA \alpha+1$. } 
\end{enumerate}
\item[3-4.]{If $S_{\alpha}\cap S_{K} \neq \phi$, $\alpha$ is removed from $C$. This is checked for all $\alpha \in C$.}
\item[3-5.]{ $C$, $\{ S_{\alpha} \}_{\alpha \in C}$, and $S_{I}=S_A-\cup_{\alpha \in C}S_{\alpha}$ are returned.}
\end{enumerate}
The entire procedure presented above implements \Reqs{LOOE-approx}{chi2}. 

A debatable point would be the values of $\theta$ and $\delta$. In most of iterative algorithms as the one in~\cite{Chambolle:04,Beck:09-1}, there is an inevitable finite error of the TV term even when it should vanish. Let us express the ``scale'' of this error as $t_{i}(\hat{\V{x}})\approx \theta_{\rm num}>0$. By construction, the threshold $\theta$ is related to this numerical error and it is appropriate to choose $\theta \approx \theta_{\rm num}$; the softening constant $\delta$ should be sufficiently larger than $\theta_{\rm num}$ because it does implement the assumed order of two limits, $\lim_{\delta \to 0}\lim_{p,q\to 0}$, in derivation of the approximation formula. Overall, the relation 
\be
\theta \approx \theta_{\rm num}\ll \delta
\ee
must be satisfied. We have numerically checked how strict this principled relation is, and found that the approximation result is not sensitive to the choice of $\theta$ as long as it is sufficiently smaller than $\delta$. Although a little more delicate points are involved in the choice of $\delta$, we have also found that in a wide range of $\delta$ the approximation result is stable and the cost-function Hessian is safely invertible. Based on these observations, in the application of our formula below, the default values are set to be $\delta=10^{-4}$ and $\theta=10^{-12}$. They are chosen according to our datasets and experimental setup: The maximum value of the non-softened TV terms is scaled as $\max_{i}t_{i}(\hat{\V{x}})\gtrsim 10^{-4}$ and the numerical precision is about $\theta_{\rm num}\approx 10^{-12}$; the former value is reflected to $\delta$ and the latter one is used in $\theta$. Coincidently, this default value of $\delta$ accords with the one in \cite{Chan:01}. The examination result of the sensitivity to $\delta$ and $\theta$ will be reported below. 

Another noteworthy point is that these procedures can be easily extended to other variants of the TV. For example, for the so-called anisotropic TV~\cite{Beck:09-1}, $T_{\mathrm{ani}}=\sum_{i}\sum_{j \in \partial i}|x_j-x_i|$, we set $F=G$ in step 4 and modify the definition of the link in step 3-1 accordingly, so as to render our formula applicable. In the case of the square TV, $T_{\mathrm{sq}}=\sum_{i}\sum_{j \in \partial i}(x_j-x_i)^2 \equiv (1/2)\V{x}^\top J\V{x}$, the formula can be significantly simpler, because this TV has no sparsifying effect and the formula of the simple $\ell_1$ case can be employed. We can employ \Req{LOOE-approx} with $\chi_{S_A S_A}=(G_{S_A S_A}+\lambda_{T}J_{S_A S_A})^{-1}$ directly, without the need for cluster enumeration.

\section{Application to super-resolution imaging}
To test the usefulness of the developed formula, 
let us apply the derived expression to the super-resolution reconstruction of astronomical images. A number of recent studies have demonstrated that sparse modeling is an effective means of reconstructing astronomical images obtained through radio interferometric observations~\cite{Wiaux:09,Li:11,Honma:14,Honma:16}. In particular, the capability of high-fidelity imaging in super-resolution regimes has been shown, which renders this technique a useful choice for the imaging of black holes with the EHT~\cite{Honma:14,Honma:16,Ikeda:16,Akiyama:17-1,Akiyama:17-2}. We adopt the same problem setting as~\cite{Honma:14,Akiyama:17-1} and demonstrate the efficacy of our approximation formula through comparison with the literally conducted 10-fold CV result. Here, $x_i$ denotes the $i$th pixel value and $A$ is (part of) the Fourier matrix. The dataset $\V{y}$ is generated through the linear process
\be
\V{y}=A\V{x}_0+\V{\xi},
\ee
where $\V{\xi}$ is a noise vector and $\V{x}_0$ is the simulated image, which we infer given $\V{y}$ and $A$.  

In this work, we use data for simulated EHT observations based on three different astronomical images, which are available as sample data for the EHT Imaging Challenge. Our datasets 1, 2, and 3 correspond to the sample datasets 1, 2, and 5, respectively, available from~\cite{ehtimagech} at July 2017. The images are reconstructed with $N=10000=100\times 100$ pixels and with 160, 250, and 100~$\rm \mu$as fields of view, which are identical to the original images of Datasets 1, 2, and 3 from the EHT Imaging Challenge, respectively. We test four values for each $\lambda_{\ell_{1}}$ and $\lambda_{T}$: $\lambda_{\ell_{1}} \in (M/2)\times \{1,10,100,1000\}$ and $\lambda_T \in (M/8)\times \{1,10,100,1000\}$. $M$ is $1910, 1910$, and $2786$, for Datasets 1--3, respectively. Later, we also use different size data from the same datasets, for checking the size dependence of the result.

Table \ref{tab:errors} shows the mean CVE values for the three datasets, determined by the 10-fold CV and by our approximation formula for varying $\lambda_{T}$.
\begin{table}[htbp]
\small
\begin{tabular}{|c||c|c|c|c|c|}
\hline
& $8\lambda_{T}/M$  &  1             & 10                    & 100          & 1000\\
\hline
\multirow{2}{*}{Dataset 1}  
& 10-fold                    & 1.101(47) & {\bf 1.090(44)} & 1.091(44) & 1.455(108) \\ 
\cline{2-6}
& Approx.                   & 1.087(35) & {\bf 1.080(35)} & 1.082(35) & 1.385(49)\\
\hline
\multirow{2}{*}{Dataset 2}  
& 10-fold                    & 1.368(91) & {\bf 1.260(55)} & 1.286(65) & 2.843(234) \\ 
\cline{2-6}
& Approx.                   & 1.180(37) & {\bf 1.157(36)} & 1.210(37) & 2.669(108)\\
\hline
\multirow{2}{*}{Dataset 3}   
& 10-fold                    & 1.026(18) & {\bf 1.020(19)} & 1.020(22) &1.235(52) \\ 
\cline{2-6}
& Approx.                   & 1.028(26) & {\bf 1.018(26)} & 1.020(27) & 1.226(40)\\
\hline
\end{tabular}
\caption{\label{tab:errors}
CVE values determined by 10-fold CV and our approximation formula against $\lambda_{T}$. $\lambda_{\ell_1}$ is fixed to the optimal value ($2\lambda_{\ell_1}/M=1$, coincidentally common to all cases). The number in brackets denotes the error bar to the last digits. The optimal values are bolded. The tuning constants $\delta$ and $\theta$ are set to be $\delta=10^{-4}$ and $\theta=10^{-12}$, respectively.}
\end{table}
 $\lambda_{\ell_1}$ is fixed to the optimal value, which is coincidently common for all datasets and satisfies $2\lambda_{\ell_1}/M=1$. It is clear that the approximate CVE values accord well with the 10-fold results, even on the error-bar scale, demonstrating that our approximation formula works very well. Note that the error bar for the approximation is given by the standard deviation of the $M$ terms in \Req{LOOE-approx} divided by $\sqrt{M-1}$. 

To directly observe the reconstruction quality, in \Rfig{Images} we display the images at all investigated parameters and the reconstructed image at the optimal $\lambda_{\ell_1}$ and $\lambda_{T}$ for Dataset $3$, as well as the associated errors plotted against $\lambda_{\ell_1}$ and $\lambda_{T}$ in \Rfig{CVE-normal}.
\begin{figure}[htbp]
\begin{center}
  \begin{tabular}{ccc}
  	\includegraphics[width=0.43\textwidth]{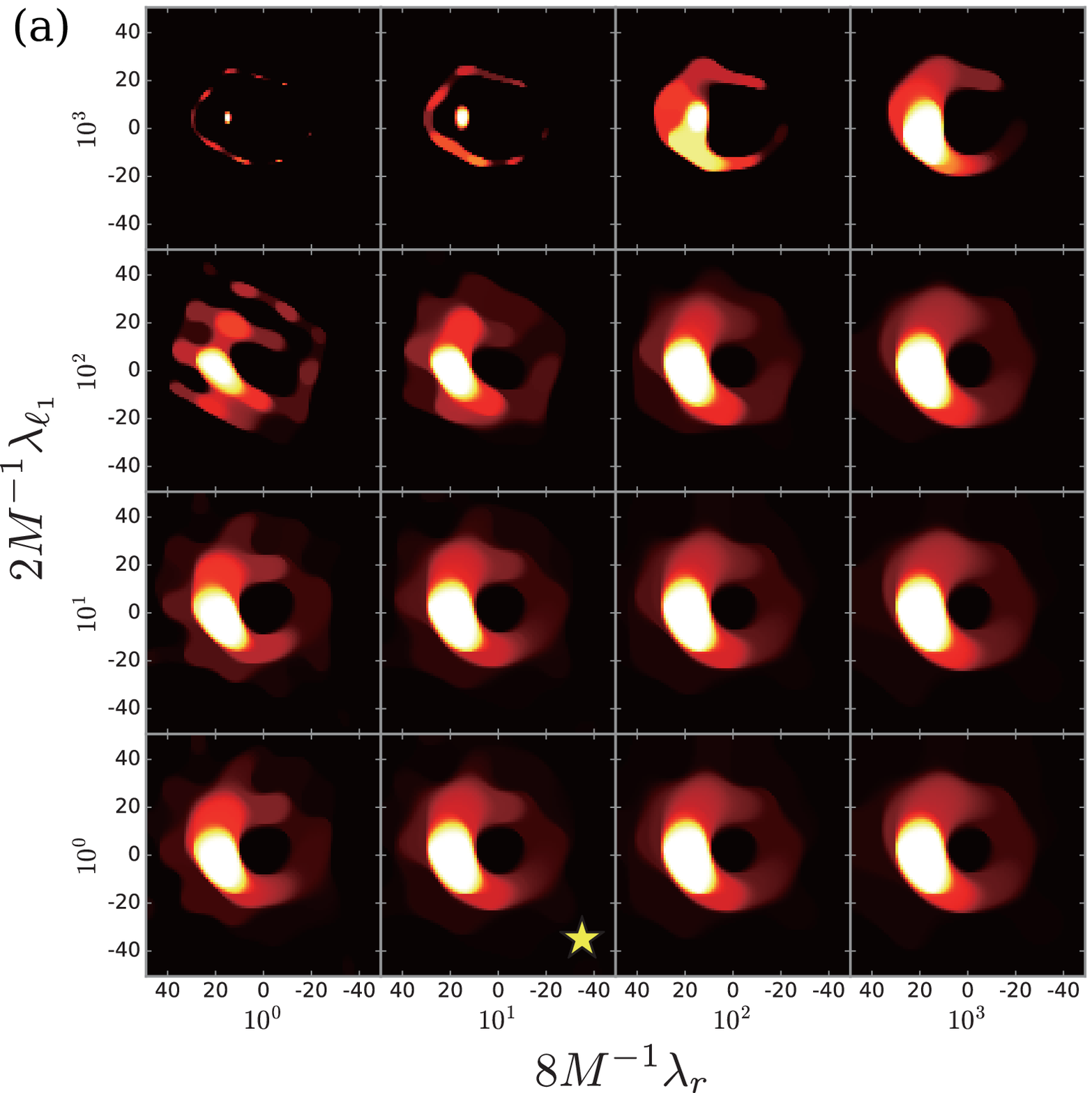} &
  	\includegraphics[width=0.44\textwidth]{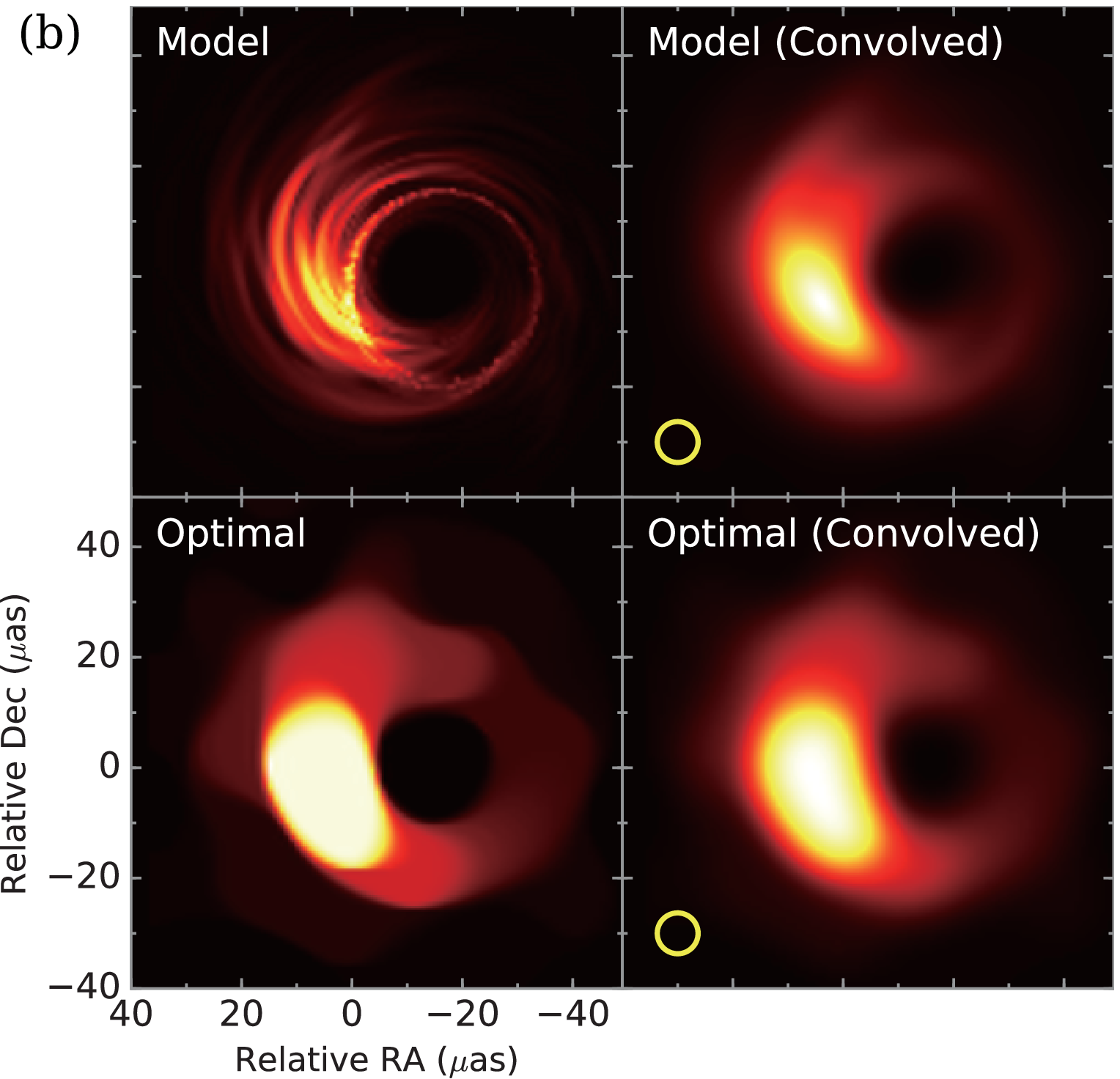} &
  \end{tabular}
  \caption{Super-resolution imaging results for Dataset 3 based on model image of supermassive black hole at center of nearby elliptical galaxy, M87. (a) Images for all investigated parameters; the star-marked panel is obtained at the optimum parameters. (b) Original images (top) and reconstructed images (bottom) at optimal parameters ($(2\lambda_{\ell_1},8\lambda_{T})/M=(1,10)$). The images are convolved with a circular Gaussian beam on the right-hand side, the full width at half maximum (FWHM) of which is 25\% of the nominal angular resolution of the EHT and corresponds to the diameters of the yellow circles. This coincides with the optimal resolution minimizing the mean square error between them. }
\Lfig{Images}
\end{center}
\end{figure}
\begin{figure}[htbp]
\begin{center}
  \begin{tabular}{ccc}
  	\includegraphics[width=0.33\textwidth]{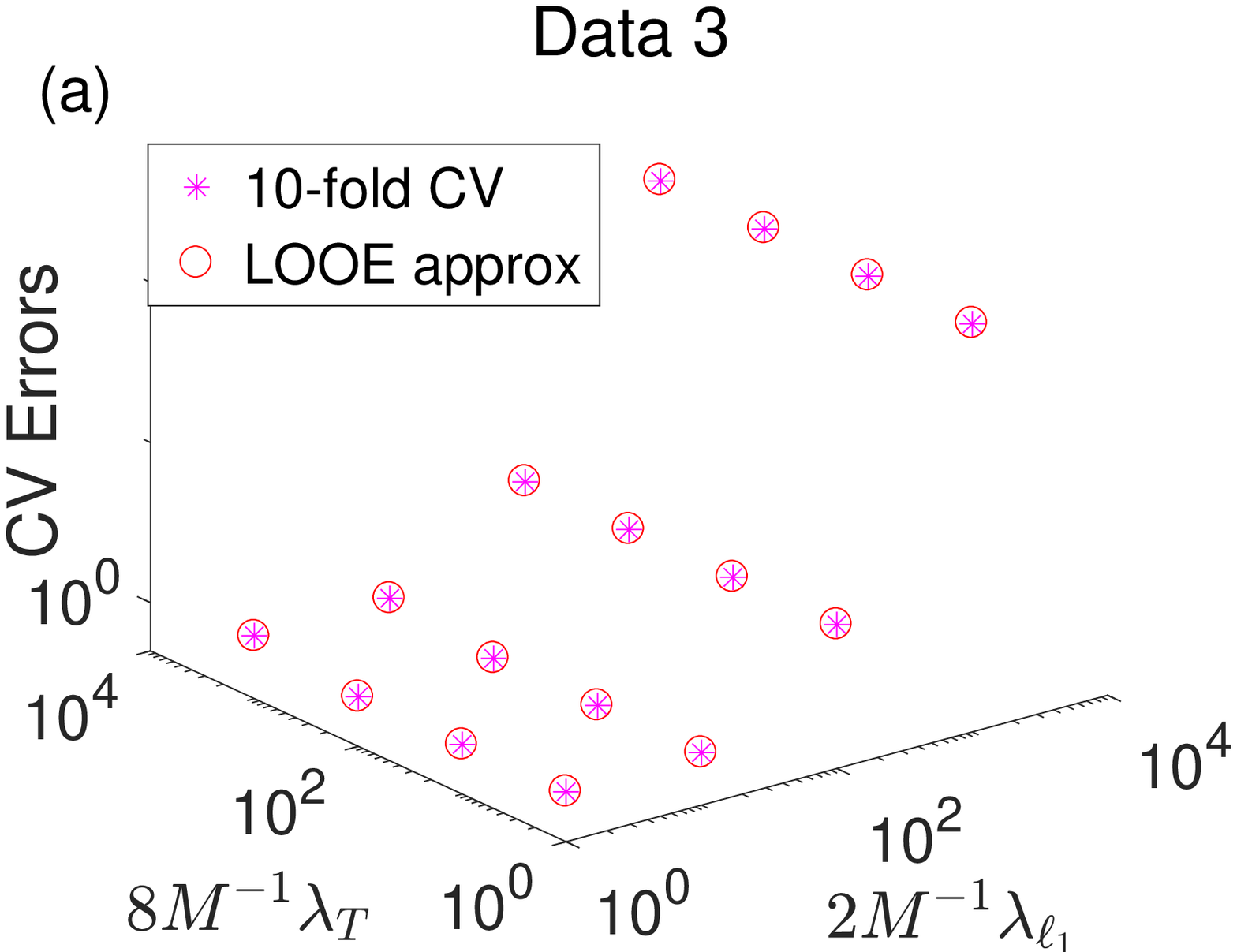} 
  	\includegraphics[width=0.33\textwidth]{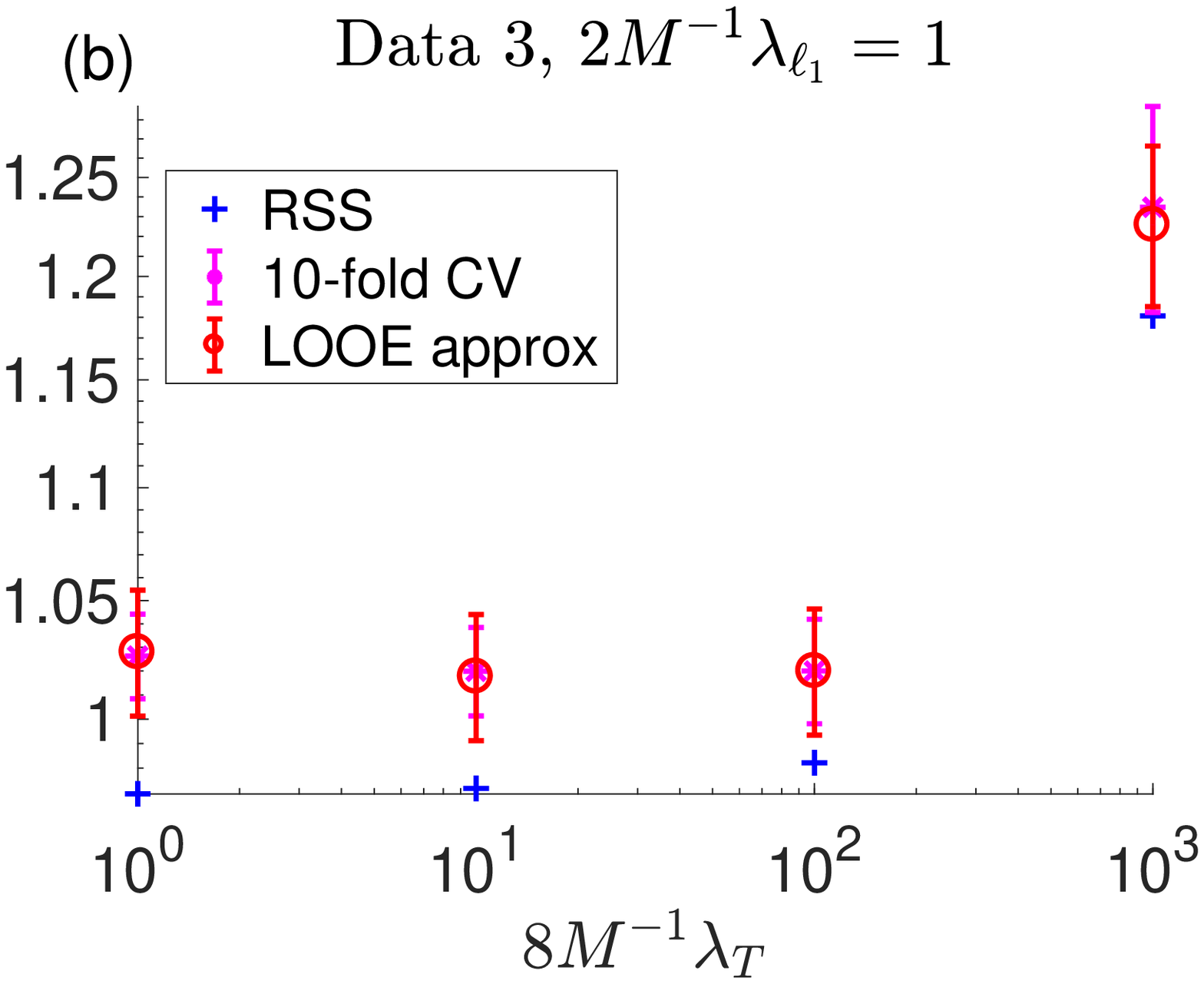} 
  	\includegraphics[width=0.33\textwidth]{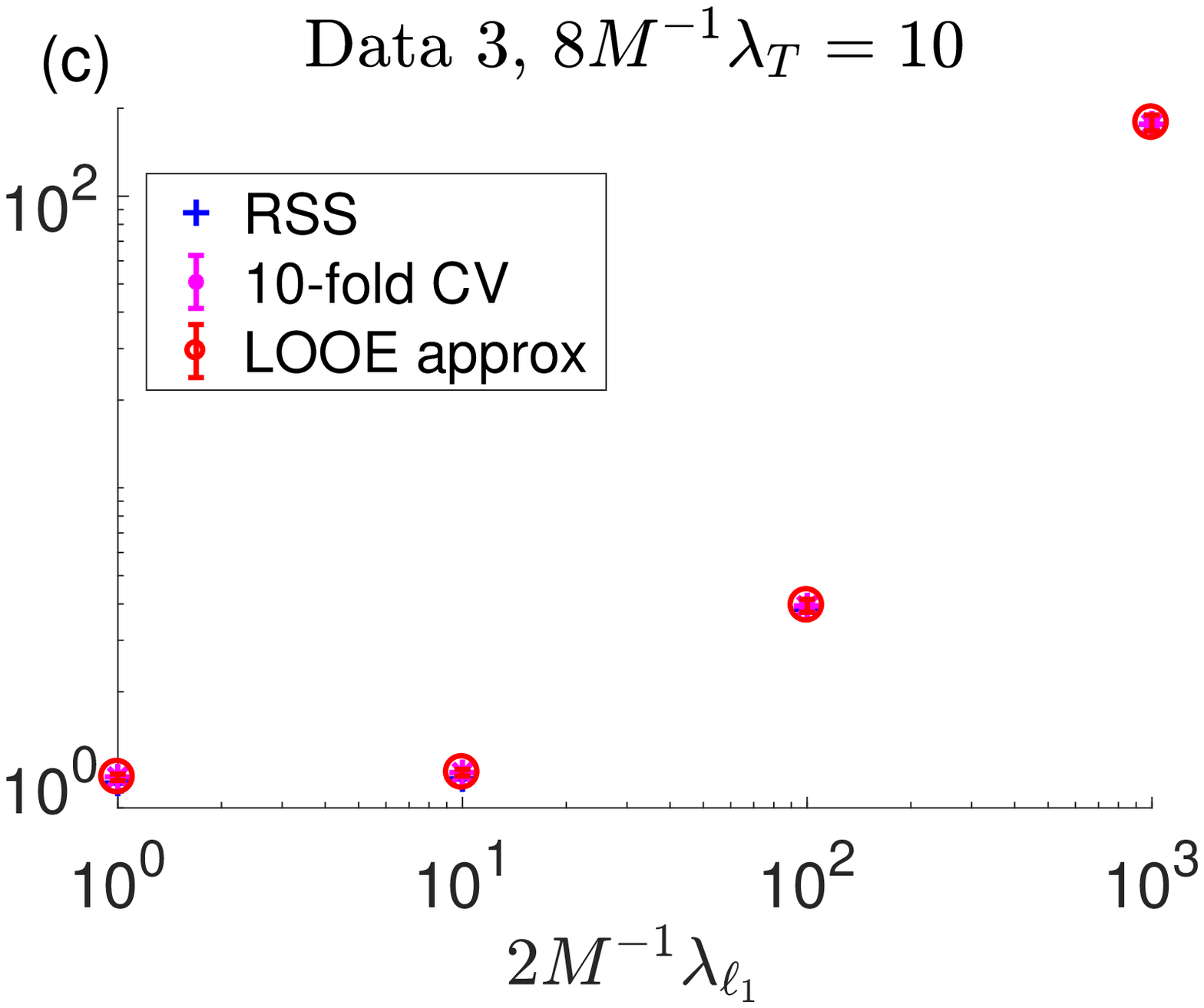} 
  \end{tabular}
  \caption{(a) 3D plot of mean CVEs against $\lambda_{\ell_1}$ and $\lambda_{T}$ without error bars. (b) Plot of mean CVEs and RSS against $\lambda_{T}$ at the optimal value of $\lambda_{\ell_1}$,  $2\lambda_{\ell_1}/M=1$.  (c) Plot of mean CVEs and RSS against $\lambda_{\ell_1}$ at the optimal value of $\lambda_T$, $8\lambda_{T}/M=10$. For (c), the RSS is overlapped with the CVEs in the symbol size. In all the cases, the agreement between the approximate LOOE and the 10-fold CVE is fairly good. The tuning constants $\delta$ and $\theta$ are set to be $\delta=10^{-4}$ and $\theta=10^{-12}$, respectively.}
\Lfig{CVE-normal}
\end{center}
\end{figure}
Again, we can see the proposed method approximates the 10-fold result well, and the reconstructed image reasonably resembles the original. The RSS is monotonic with respect to the changes of $\lambda_{l_1}$ and $\lambda_T$ but the approximate LOOE is not, which implies that the LOOE factor computed through \Req{chi2} appropriately reflects the effect of the penalty terms.

Next, we check the sensitivity of the approximate result to the tuning constants $\delta$ and $\theta$. In \Rfig{CVE-sensitivity}, the approximate LOOEs at the optimal $\lambda_{\ell_1}$ are plotted against $\lambda_T$ when changing $\delta$ (left) and $\theta$ (right).
\begin{figure}[htbp]
\begin{center}
  \begin{tabular}{ccc}
  	\includegraphics[width=0.42\textwidth]{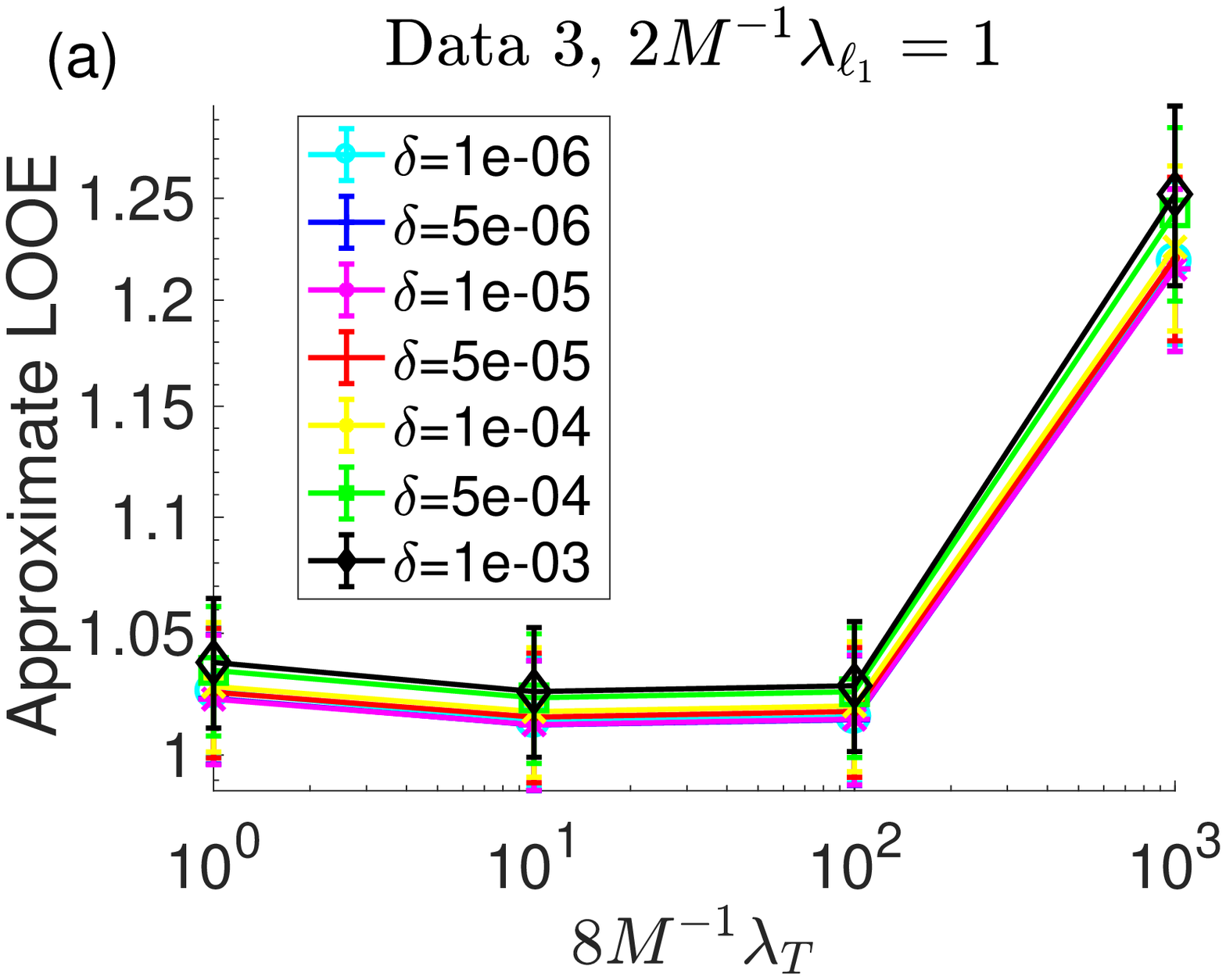} 
  	\includegraphics[width=0.42\textwidth]{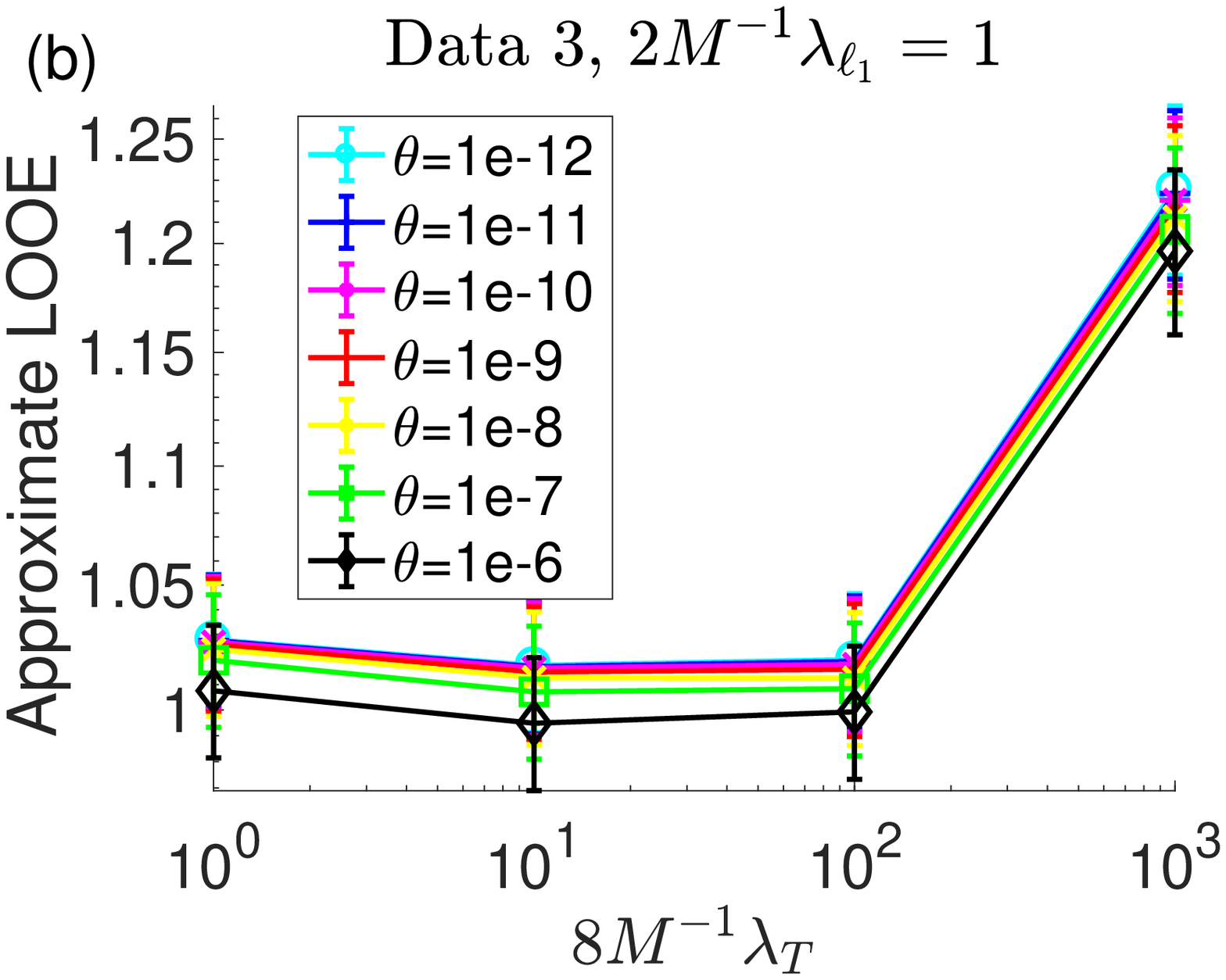} 
  \end{tabular}
  \caption{Comparative plots of mean approximate LOOEs against $\lambda_{T}$ at $2M^{-1}\lambda_{\ell_1}=1$ when (a) $\delta$ changes as $10^{-6}$--$10^{-3}$ with fixed $\theta=10^{-12}$; (b) $\theta$ changes as $10^{-12}$--$10^{-6}$ with fixed $\delta=10^{-4}$. They show that the LOOE curves are rather stable against the choice of the tuning constants. }
\Lfig{CVE-sensitivity}
\end{center}
\end{figure}
This indicates that the approximate LOOEs are stable against the change of both $\delta$ and $\theta$. Hence, we may choose these values rather arbitrarily. This is a good news because tuning them makes the problem more numerically amenable: Enlarging $\delta$ makes the computation of the Hessian inversion more numerically stable; increasing $\theta$ lowers the effective degrees of freedom. The second property associated with $\theta$ is really beneficial when treating a large-size dataset, because it can downsize the Hessian and reduce the cost for computing its matrix inversion. In Table \ref{tab:DF}, the values of the effective degrees of freedom are given when changing $\theta$. The reduction of the degree of freedom at large (yet small enough compared to $\delta=10^{-4}$) $\theta$ is significant, which encourages us to apply the proposed formula to larger-size datasets.  
\begin{table}[htbp]
\begin{tabular}{|c||c|c|c|c|c|c|c|}
\hline
 $ \theta $                & 1e-12 & 1e-11 & 1e-10 & 1e-09 & 1e-08 & 1e-07 & 1e-06 \\
\hline
 $ |\tilde{S}_{I+Z}| $ &  5733  & 5524  & 5243  & 4814  & 4112  & 2922  & 1408 \\
\hline
\end{tabular}
\caption{\label{tab:DF}
The effective degrees of freedom $|\tilde{S}_{I+Z}|$, the number of clusters + the number of isolated variables, against $\theta$ for Dataset 3 at $\delta=10^{-4}$ and the optimal parameters $(2\lambda_{\ell_1},8\lambda_{T})/M=(1,10)$.} 
\end{table}

Finally, let us see the data-size dependence of the approximation accuracy and of the computational cost for solving \Req{Problem} and for obtaining the approximate LOOE from the solution. The data analyzed here is an identical simulated image of black hole expressed with different number of pixels. When solving \Req{Problem}, we used Intel(R) Core(TM) i7-5820K CPU of 3.30GHz with 6 cores for $N=50^2=2500$ and Intel(R) Xeon(R) CPU E5-2699 v3 of 2.30GHz with 36 cores for $N=100^2$ and $150^2$, and employed an algorithm called ``MFISTA'' proposed in~\cite{Chambolle:04,Beck:09-1}. Meanwhile, we used a laptop of a 1.7 GHz Intel Core i7 with two CPUs for evaluating the approximate LOOE. Hence the comparison is not fair and unfavorable to the approximation formula.
\begin{figure}[htbp]
\begin{center}
  \begin{tabular}{ccc}
  	\includegraphics[width=0.42\textwidth]{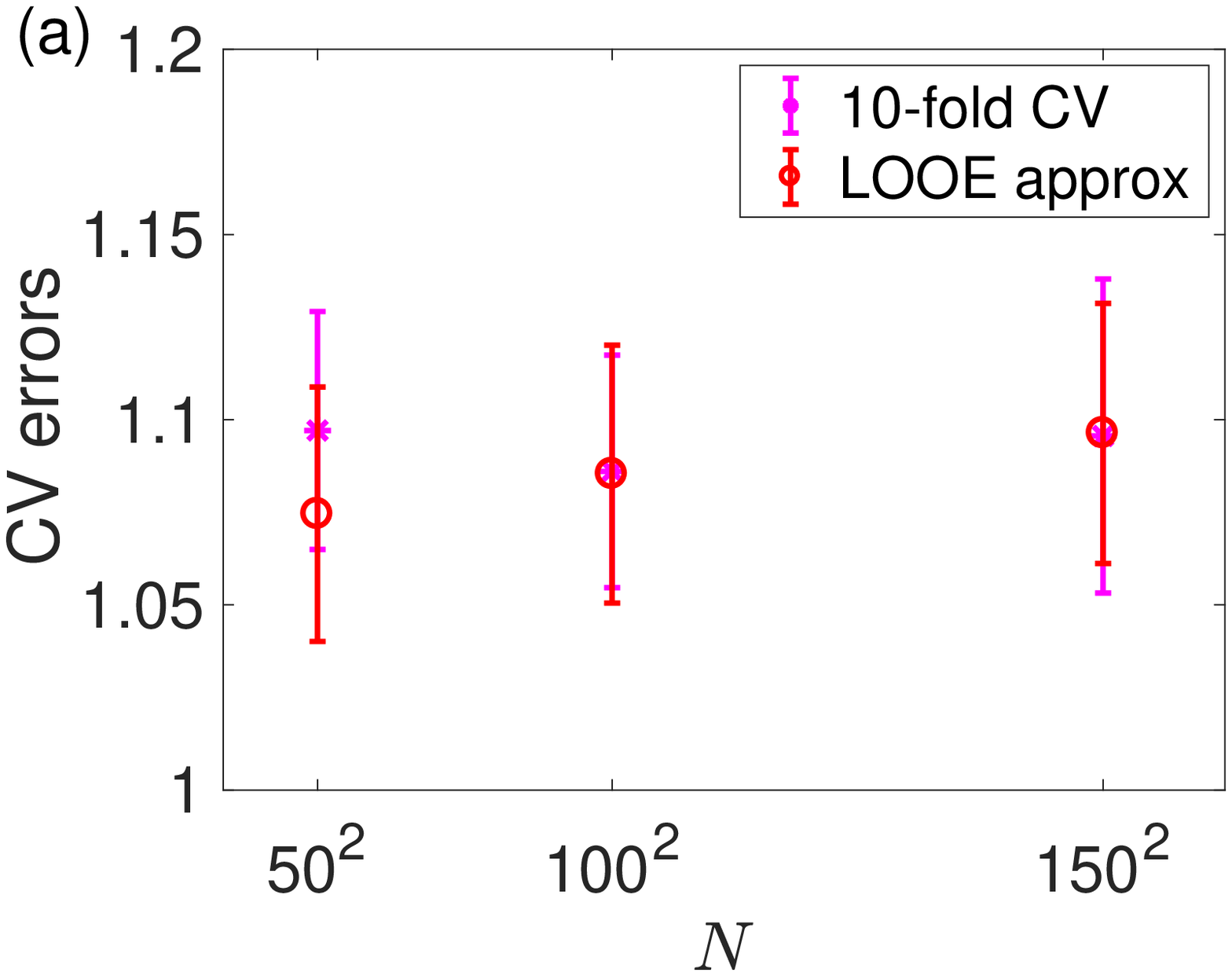} 
  	\includegraphics[width=0.42\textwidth]{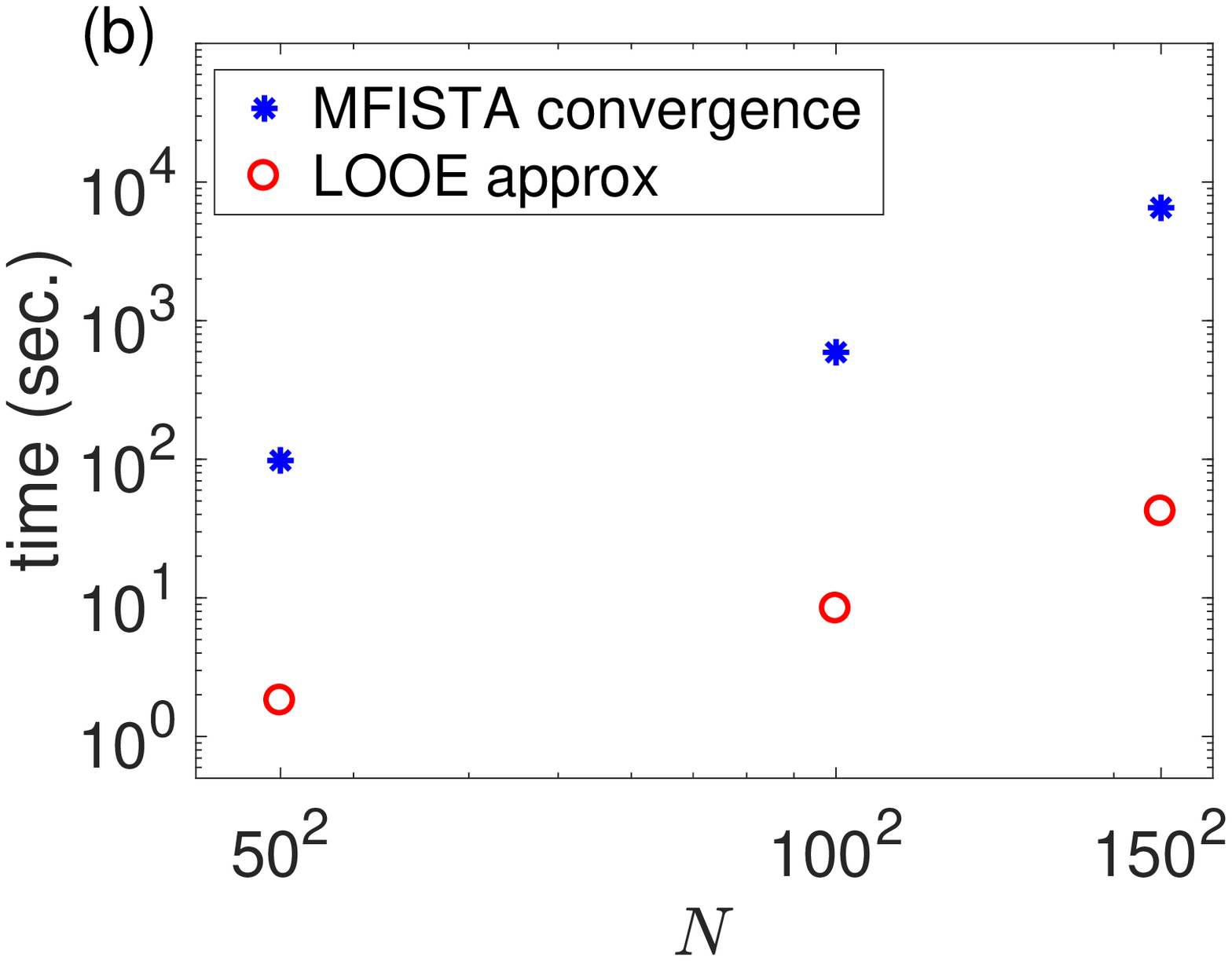} 
  \end{tabular}
  \caption{(a) Plot of mean CVEs at optimal parameters of different sizes. (b) Log-log plot of the computational times for solving the optimization problem \NReq{Problem} and for obtaining the approximate value of CVE against the size of datasets. } 
\Lfig{CVE-size}
\end{center}
\end{figure}
The left panel indicates that the approximation accuracy becomes better for larger sizes. This is reasonable because the perturbation we have employed should have better accuracy as the model and data become larger, though the accuracy at $N=50^2=2500$ is already good. The right panel clearly shows the advantage of the developed formula: The actual computational time of the approximate LOOE is significantly shorter than that of the algorithm convergence for solving \Req{Problem} in the investigated range of system sizes, even under the unfair comparison mentioned above. However, this advantage will be less prominent if the model becomes very large: Our approximation formula needs the Hessian inversion whose computational cost is scaled as $O( (|C|+|S_{I}|)^3 )\approx O(N^3)$, while MFISTA requires the cost of $O(N^2)$ as long as the number of steps to convergence is constant against $N$. The crossover size at which these two computational costs become comparable is roughly estimated as $N_{\times}\approx 10^6$, though such crossover tendency cannot be seen yet from \Rfig{CVE-size}. For such large systems, a new fundamental solution should be tailored to resolve the computational-cost problem, though tuning $\theta$ to a large value in the present method can still be a good first aide. 

\section{Conclusion} 
In this paper, we have developed an approximation formula for the CVE of a sparse linear regression penalized by $\ell_1$ and TV terms, and demonstrated its usefulness in the reconstruction of simulated black hole images. Our derivation is based on the perturbation assuming the small difference between the full and leave-one-out solutions. This assumption will not be fulfilled for some specific cases, i.e. when the measurement matrix is sparse. However, for most of dense measurement matrices, such as the Fourier matrix discussed in this paper, our assumption will be reasonably satisfied.  Hence we expect the range of application of our formula is wide enough and we would like encourage the readers to use this formula in their own work.
It is also straightforward to generalize the developed formula to other types of TV, and two examples of the generalization for the anisotropic and square TVs have been explained.

The key concept of our formula, perturbation between the LOO and full systems, is very general and can be applied to more general statistical models and inference frameworks~\cite{Kabashima:16}. The development of practical formulas for those cases will facilitate higher levels of modeling and computation. 

A Matlab package implementing the approximation formula is available from~\cite{Github:Obuchi}.

\section{Acknowledgements} 
We would like to express our sincere gratitude to Mareki Honma and Fumie Tazaki for their helpful discussions. We thank Katherine L. Bouman for preparing the EHT Imaging Challenge website~\cite{ehtimagech,Bouman:16}. We also thank Andrew Chael and Lindy Blackburn for writing a simulation software to produce sample data sets~\cite{Chael:16}. 


\end{document}